\documentclass[prl,twocolumn,aps,superscriptaddress,showpacs,floatfix]{revtex4}
\usepackage{amsmath}
\usepackage{graphicx}
\begin{document}
\bibliographystyle{apsrev}

\title{A mechanism for phase separation in copper oxide superconductors}

\author{A.O. Sboychakov}
\affiliation{Frontier Research System, The Institute of Physical
and Chemical Research (RIKEN), Wako-shi, Saitama, 351-0198, Japan}
\affiliation{Institute for Theoretical and Applied Electrodynamics
Russian Academy of Sciences, 125412 Moscow, Russia}

\author{Sergey Savel'ev}
\affiliation{Frontier Research System, The Institute of Physical
and Chemical Research (RIKEN), Wako-shi, Saitama, 351-0198, Japan}
\affiliation{Department of Physics, Loughborough University,
Loughborough LE11 3TU, United Kingdom}

\author{A.L. Rakhmanov}
\affiliation{Frontier Research System, The Institute of Physical
and Chemical Research (RIKEN), Wako-shi, Saitama, 351-0198, Japan}
\affiliation{Institute for Theoretical and Applied Electrodynamics
Russian Academy of Sciences, 125412 Moscow, Russia}
\affiliation{Department of Physics, Loughborough University,
Loughborough LE11 3TU, United Kingdom}

\author{K.I. Kugel}
\affiliation{Institute for Theoretical and Applied Electrodynamics
Russian Academy of Sciences, 125412 Moscow, Russia}
\affiliation{Department of Physics, Loughborough University,
Loughborough LE11 3TU, United Kingdom}

\author{Franco Nori}
\affiliation{Frontier Research System, The Institute of Physical
and Chemical Research (RIKEN), Wako-shi, Saitama, 351-0198, Japan}
\affiliation{Center for Theoretical Physics, CSCS, Department of
Physics, University of Michigan, Ann Arbor, MI 48109-1040, USA}

\begin{abstract}
A two-band Hubbard model is used to describe the band structure
and phase separation (PS) in multiband superconductors, especially
in cuprates. We predict a large peak in the density of states at
the Fermi level in the case of optimum doping, corresponding to
the minimum energy difference between the centers of two hole
bands. For strong interband hybridization, a metal-insulator
transition occurs near this optimum doping level. We suggest a
mechanism of PS related to the redistribution of holes between two
Hubbard bands rather than to the usual antiferromagnetic
correlations. We show that the critical superconducting
temperature $T_c$ can be about its maximum value within a wide
range of doping levels due to PS.
\end{abstract} 

\date{\today}

\maketitle

Nanoscale spatial variations in the electronic characteristics of
different high-$T_c$ superconductors have been commonly observed
in the form of stripes~\cite{Dag}, granular droplet-like
structures~\cite{droplets}, four-unit-cell periodic (checkerboard)
patterns~\cite{4x4}, or even more intricate
arrangements~\cite{else}. Several theoretical models also suggest
inhomogeneous electron structure in superconducting cuprates and
related materials~\cite{theor, theor1,theor2,theor3}. Moreover,
the phenomenon of self-organized electron inhomogeneity or phase
separation (PS) is common to many strongly correlated electron
systems~\cite{Nag,Dag}. For example, PS in the form of droplets,
stripes, and checkerboard patterns has been observed in doped
magnetic oxides (manganites)~\cite{Dag}. Many theoretical
approaches~\cite{theor2,theor3,Nag,Dag} explain PS as a result of
the competition between electron localization, due to
antiferromagnetic correlations, and delocalization in the
non-magnetic (or ferromagnetic) regions.

Here we consider an alternative mechanism (\textit{without}
antiferromagnetic correlations) for the PS arising in a Hubbard
model with two (or more) bands. Even though this model is commonly
used to describe high-$T_c$ superconductors~\cite{book}, PS has
not been studied using such an approach. The two-band Hubbard
model allows us to describe PS of the droplet type (including the
droplet size) observed in experiments (see,
e.g.,~\cite{droplets,else}). In contrast to the usual mechanism,
based on antiferromagnetic correlations, here we show that our
proposed mechanism of PS gives rise to a density of states (DOS)
at the Fermi level, which corresponds to the optimum doping in one
of the phases. This results in a weak variance of the critical
temperature within a broad doping range, as observed in
experiments. The number of charge carriers in superconducting
cuprates is not directly determined by the doping level and can
depend on temperature~\cite{book,GT}. This fundamental problem can
be also understood in the framework of two-band models~\cite{GT}.

\textit{Two-band Hubbard model.}--- The usual model used to
describe superconducting cuprates is based on the Hubbard
Hamiltonian with three bands: Cu(3$d$), O(2$p_x$), and
O(2$p_y$)~\cite{book}. In earlier studies, this Hamiltonian was
reduced to an effective single-band Hubbard Hamiltonian (see,
e.g.,~\cite{book,ZR}). However, recent computations (see,
e.g.,~\cite{Good,Hoz}) show that two distinguishable Cu-O bond
lengths in the CuO$_6$ octahedron, and the direct tunnelling of
holes between the oxygen atoms lead to the effective two-band
Hubbard Hamiltonian~\cite{Hoz}:
\begin{eqnarray}\label{H}
\nonumber{\cal
H}=-\!\!\!\sum_{\langle\mathbf{nm}\rangle\alpha\beta\sigma}t^{\alpha\beta}a^{\dag}_{\mathbf{n}\alpha\sigma}a_{\mathbf{m}\beta\sigma}
-\sum_{\mathbf{n}\alpha\sigma}(\mu+\epsilon_\alpha)n_{\mathbf{n}\alpha\sigma}
\\+\frac{1}{2}\sum_{\mathbf{n}\alpha,\sigma}U^{\alpha}n_{\mathbf{n}\alpha\sigma}n_{\mathbf{n}\alpha\bar{\sigma}}
+\frac{U'}{2}\sum_{\mathbf{n}\alpha,\sigma\sigma'}n_{\mathbf{n}\alpha\sigma}n_{\mathbf{n}\bar{\alpha}\sigma'}\,.
\end{eqnarray}
Here, $a^{\dag}_{\mathbf{n}\alpha\sigma}$ and
$a_{\mathbf{n}\alpha\sigma}$ are the creation and annihilation
operators for holes in the state $\alpha=\{p,\,d\}$ at site
$\mathbf{n}$ with spin projection $\sigma$
($\bar{\alpha},\bar{\sigma}$ denote not-$\alpha$ and
not-$\sigma$), the symbol $\langle\dots\rangle$ denotes a
summation over the nearest sites, $\mu$ is the chemical potential,
$\epsilon_d$ is the energy difference between the centers of the
$d$ and $p$ bands and $\epsilon_p=0$. The first term in
Eq.~\eqref{H} is the kinetic energy of the conduction holes, the
second term is due to the chemical potential and the shift between
the centers of $d$ and $p$ bands. The last two terms correspond,
respectively, to the intra- and interband on-site Coulomb
repulsion. In agreement to Ref.~\onlinecite{Hoz}, we assume that
the Coulomb interaction is strong enough, that is
$U^{\alpha},\,U'\gg t^{\alpha\beta},\,\epsilon_d$.
Applying the mean field (so-called Hubbard I) approximation, $\langle\hat{T}a_{\mathbf{m}\alpha\sigma}(t)n_{\mathbf{n}\beta\sigma'}(t)%
a^{\dag}_{\mathbf{n}_0\alpha\sigma}(t_0)\rangle\to\langle%
n_{\mathbf{n}\beta\sigma'}\rangle\langle\hat{T}%
a_{\mathbf{m}\alpha\sigma}(t)a^{\dag}_{\mathbf{n}_0\alpha\sigma}(t_0)\rangle$
($\hat{T}$ is time-ordering operator), for Hamiltonian~\eqref{H},
we derive the relationship
\begin{equation}\label{EqGab}
\!\!\!\left(\omega\!+\!\mu\!+\!\epsilon^{\alpha}\right)G_{\alpha\beta}(\omega,\mathbf{k})\!=\!%
g_{\alpha}\!\!\left(\!\!1\!+\!\sum_{\gamma}\!\varepsilon^{\alpha\gamma}(\mathbf{k})G_{\gamma\beta}(\omega,\mathbf{k})\!\!\right)\,,
\end{equation}
for the one-particle Green's
functions $G_{\alpha\beta,\sigma\sigma'}(\mathbf{n}-\mathbf{n}_0,\,t-t_0)=%
-i\langle\hat{T}a_{\mathbf{n}\alpha\sigma}(t)a^{\dag}_{\mathbf{n}_0\beta\sigma'}(t_0)\rangle$
in the frequency-momentum ($\omega,\textbf{k}$) representation;
where $g_\alpha=1-n_{\bar{\alpha}}-n_\alpha/2$. The form of the
function $\varepsilon^{\alpha\beta}(\mathbf{k})$ depends on the
symmetry of the crystal lattice. We analyze here a simple cubic
lattice. In this case, we obtain
$\varepsilon^{\alpha\beta}(\mathbf{k})=w^{\alpha\beta}\zeta(\mathbf{k})$,
$w^{\alpha\beta}=zt^{\alpha\beta}$, and $\zeta({\bf
k})=-\left[\cos (k_1d)+\cos(k_2d)+\cos(k_3d)\right]/3$, where $d$
is the lattice constant. Below we consider a purely paramagnetic
state, that is,
$n_{\alpha\uparrow}=n_{\alpha\downarrow}=n_\alpha/2$, and neglect
the values $\langle a^{\dag}_\alpha a_{\bar{\alpha}}\rangle$.
Note, however, that the latter assumption does not affect much the
obtained results~\cite{hermit}.

Relationship~\eqref{EqGab} is the linear set of the equations for
$G_{\alpha\beta}$, which can be easily solved. The calculated
Green's functions determine the DOS and the energy of the system.
Following this approach, we derive the DOS,
$\rho_{\alpha\beta}(E)=-\pi^{-1}\int
d^3\mathbf{k}/(2\pi)^3\text{Im}[G_{\alpha\beta}(\omega+i0,\mathbf{k})]|_{\omega+\mu=E}$,
in the form
\begin{equation}\label{rhoAB}
\rho_{\alpha\beta}(E)=\sqrt{g_{\alpha}g_{\beta}}\sum_{j=\pm1}\int\frac{d^3\mathbf{k}}{(2\pi)^3}v^j_{\alpha}v^j_{\beta}%
\;\delta\left(E-\bar{\varepsilon}_j(\mathbf{k})\right)\,,
\end{equation}
where $\bar{\varepsilon}_j(\zeta(\textbf{k}))$ and
$v^j_{\alpha}(\zeta(\textbf{k}))$ are the eigenvalues and
eigenvectors of the matrix
$\bar{\varepsilon}^{\alpha\beta}(\zeta)=\sqrt{g_{\alpha}g_{\beta}}\;w^{\alpha\beta}\zeta-\delta_{\alpha\beta}\epsilon^{\alpha}$.
Solving this eigenvalue problem, we obtain the energy spectrum
$\bar{\varepsilon}_j(\textbf{k})$ of holes in two new bands
(labelled by $j=\pm 1$) and coefficients $v^j_{\alpha}$
determining the transformation from $p$ and $d$ holes to the
quasiparticles in these bands
\begin{eqnarray}\label{es}
\nonumber
\bar{\varepsilon}_j(\zeta(\textbf{k}))&=&\frac12\Biggl\{(\bar{w}^{aa}+\bar{w}^{bb})\zeta-\epsilon\\
&-&j\sqrt{[(\bar{w}^{aa}-\bar{w}^{bb})\zeta+\epsilon]^2+4(\bar{w}^{ab})^2}\Biggr\}\,,
\end{eqnarray}
\begin{equation}\label{vs}
v^j(\zeta)=\frac{1}{\sqrt{[\bar{w}^{aa}\zeta-\bar{\varepsilon}_j]^2+(\bar{w}^{ab})^2\zeta^2}}\left(\begin{array}{c}-\bar{w}^{ab}\zeta\\
\bar{w}^{aa}\zeta-\bar{\varepsilon}_j,
\end{array}
\right)
\end{equation}
where
$\bar{w}^{\alpha\beta}=\sqrt{g_{\alpha}g_{\beta}}w^{\alpha\beta}$.
In contrast to $p$ and $d$ holes with short lifetime due to
interband transitions, new quasiparticles with spectrum \eqref{es}
have a longer lifetime and are scattered by, e.g., phonons and
impurities. The transformation performed above is similar to the
well-known Zhang-Rice derivation of the single-band effective
Hamiltonian if $t_{pp}=0$~\cite{ZR}. We denote the lower band as
$j=1$ and the upper one as $j=-1$. We can write
$\rho_{\alpha\beta}$ using the dimensionless DOS
$\rho_0(E')=\int\delta(E'-\zeta(\textbf{k}))d^3\textbf{k}/(2\pi)^3$
of uncorrelated electrons,
\begin{equation}\label{rhoAB0}
\rho_{\alpha\beta}(E)\!=\!\sqrt{g_{\alpha}g_{\beta}}\!\!\sum_{j=\pm1}\!\!v^j_{\alpha}\!(\zeta)v^j_{\beta}(\zeta)%
\left|\frac{\partial\varepsilon_j}{\partial\zeta}\right|^{-1}\!\!\!\!\!\rho_0(\zeta)|_{\zeta=\bar{\zeta}_j(E)}\,,
\end{equation}
where $\bar{\zeta}_j(E)$ is the inverse function of
$\bar{\varepsilon}_j(E)$. The number of electrons in the state
$\alpha$ is
\begin{equation}\label{nab}
n_{\alpha}=2\int_{\mu_{\text{min}}}^{\mu}dE\;\rho_{\alpha\alpha}(E,n_{a},n_{b})\,,
\end{equation}
where $\mu_{\text{min}}=\bar{\varepsilon}_{1}(\zeta=-1)$ (note
that $\rho_{\alpha\beta}$ depends on $n_{\alpha}$ through the
functions $g_{\alpha}$). The total number of charge carriers per
site is $n=\sum_{\alpha}n_{\alpha}(\mu)$. This equality, along
with Eqs.~(\ref{es})--(\ref{nab}), form a set of equations for the
calculation of $n_{\alpha}$ and $\mu$. Performing a transformation
of the operators $a_\alpha$ employing the coefficients \eqref{vs},
we can find the Green's functions corresponding to the band
spectrum \eqref{es} and derive the DOS, $\rho_j$ for two $j$-bands
\begin{equation}\label{rhoS0}
\rho_j(E)=\left[\sum_{\alpha}g_{\alpha}[v^j_{\alpha}(\zeta)]^2\right]%
\left|\frac{\partial\varepsilon_j(\zeta)}{\partial\zeta}\right|^{-1}\!\!\!\!\!\rho_0(\zeta)|_{\zeta=\bar{\zeta}_{j}(E)}\,.
\end{equation}

\begin{figure}[btp]
\begin{center}
\includegraphics*[width=7.7 cm]{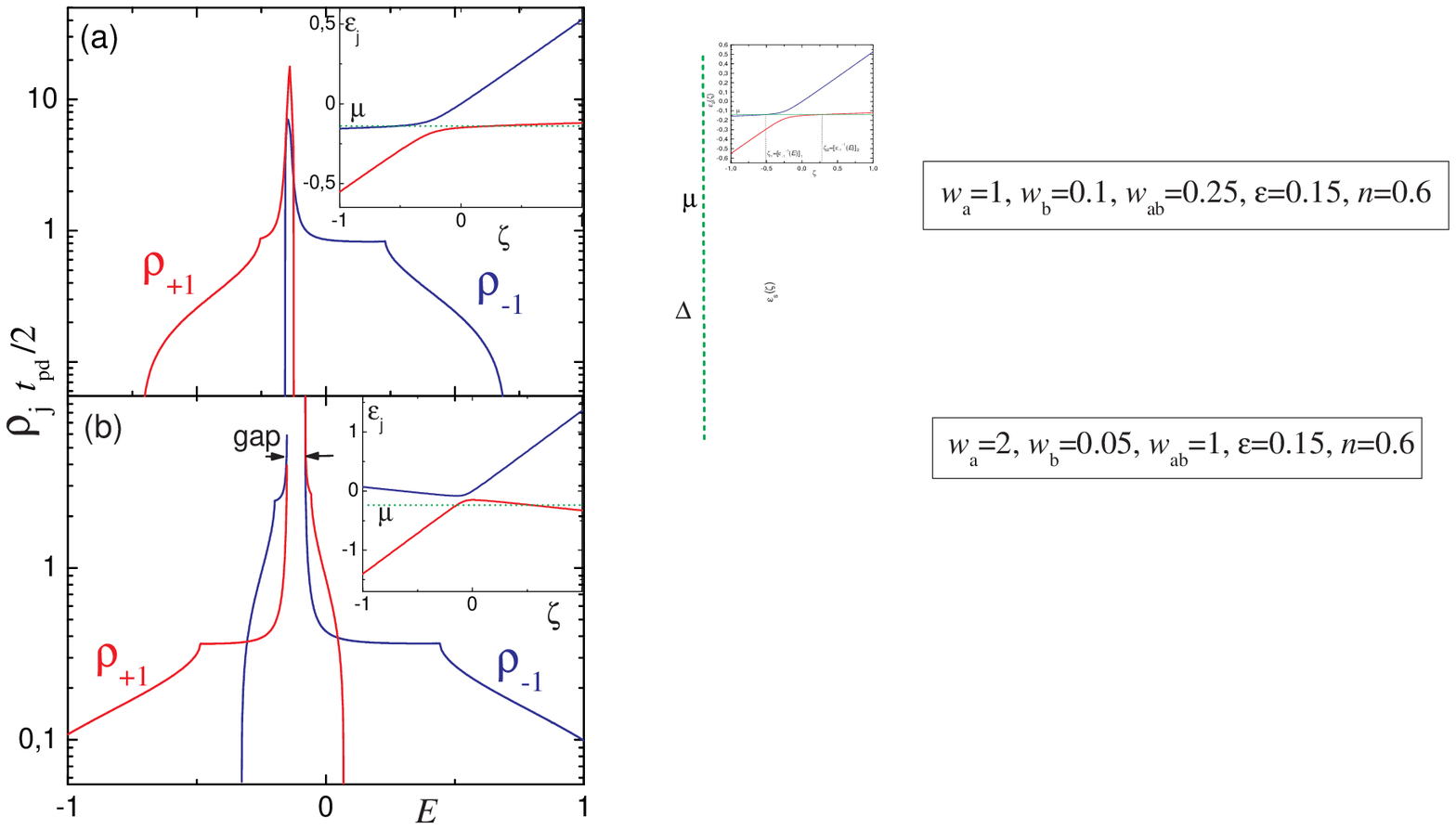}
\end{center} \caption{ (Color online) Density of states $\rho_{\pm
1}$ versus energy $E$ for quasiparticles (holes) located in bands
$j=\pm 1$ for two different types of spectra, shown in the
corresponding insets. The gapless spectrum shown in the inset (a)
was calculated at $n=0.6$, $\varepsilon_d=0.15$~eV, $w_{pp}=1$~eV,
$w_{dd}=0.1$~eV, and relatively small interband hybridization
$w_{pd}=0.25$~eV. The gapped spectrum in the inset (b) was
calculated for the same parameters but for stronger interband
hybridization $w_{pd}=1$~eV. In case (b), the transition to the
insulating state occurs at a certain doping level for which the
chemical potential $\mu$ (shown by the green dotted line) is
located inside the gap. In both cases, a large peak in the DOS is
observed at energies corresponding to the anticrossing of the
bands, where a significant flattening of the Fermi surface (see
insets) takes place. The quasiparticle energy spectra are shown in
the insets as a function of the variable $\zeta$, since
$\varepsilon_j$ depends on the crystal momentum $\textbf{k}$ only
through $\zeta(\textbf{k})$. } \end{figure}

\begin{figure}[btp]
\begin{center}
\includegraphics*[width=7.7 cm]{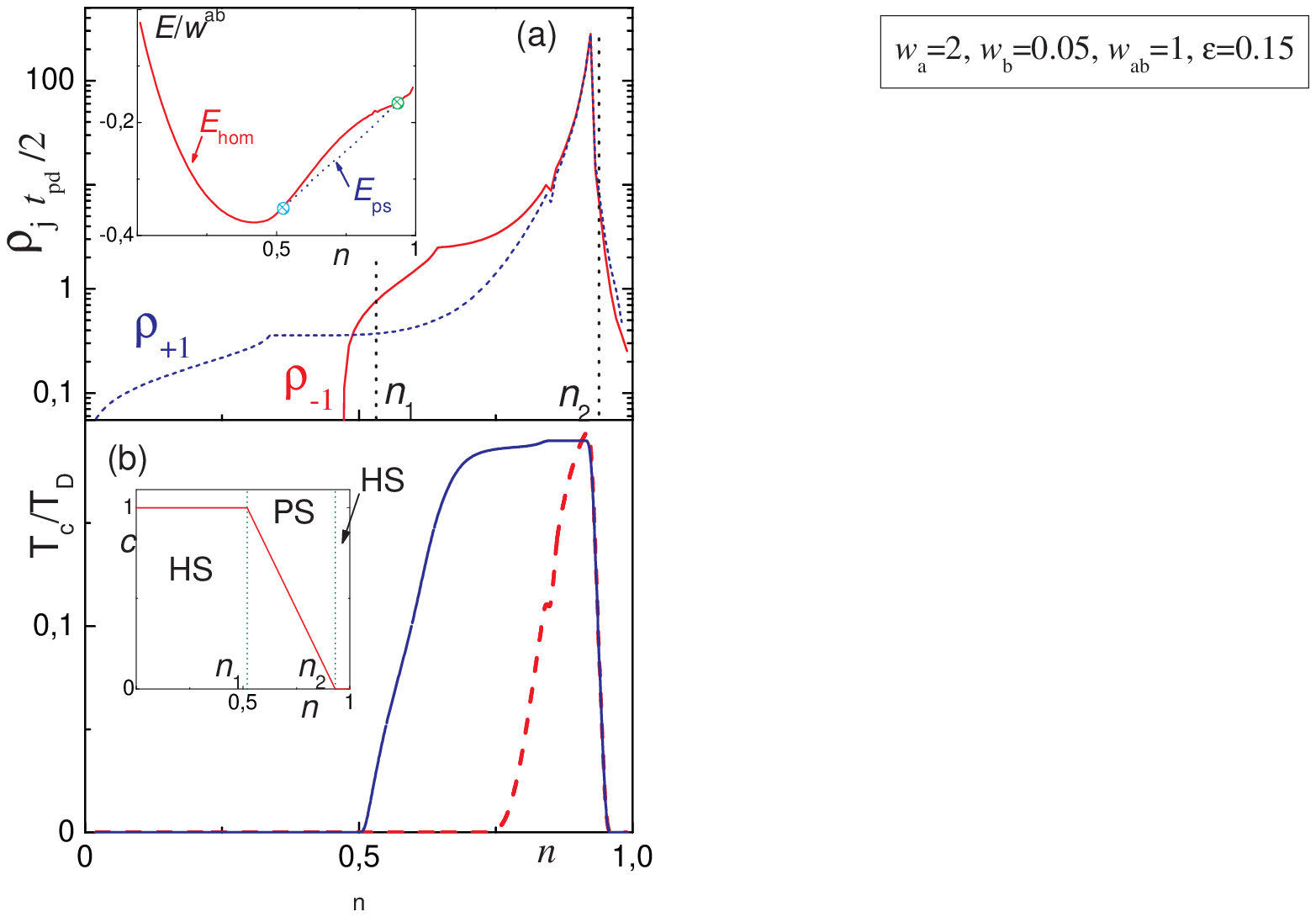}
\end{center} \caption{ (Color online) (a) The density of states at
the Fermi level versus doping in the two-band Hubbard model. Inset
in (a): energy of the homogeneous (red solid line) and the phase
separated (blue dashed line) states. The dependence of the energy
in the homogeneous state, $E_{\textrm{hom}}$, on $n$ has a
negative curvature if $n_1<n<n_2$. In this range of doping, the PS
state becomes more favorable since its energy, $E_{\textrm{ps}}$,
is lower than $E_{\textrm{hom}}$ (see inset in (a)). The values
$n_{1,2}$ are indicated in the main panel of (a) by black vertical
dotted lines. The point $n_2$ is near the peak in the DOS. (b) The
critical temperature, $T_c$, of the superconducting transition
versus doping level $n$ for the homogeneous (red dashed line) and
PS (blue solid line) states calculated using Eq.~\eqref{Tc}. The
$T_c(n)$ of the homogeneous state decreases fast when the doping
$n$ deviates from its optimum value about $n_2$. In contrast to
this, $T_c(n)$ in the PS state is a broad function and exhibits a
plateau within the $n_1<n<n_2$ range. Inset in (b) demonstrates
the dependence of the concentration $c$ of the phase with lower
hole content versus $n$. The regions of homogeneous (HS) and PS
states are indicated in the inset. Here we use the following
parameters: $\varepsilon_d=0.2$~eV, $w_{pp}=1$~eV,
$w_{dd}=0.3$~eV, and $w_{pd}=0.7$~eV. }
\end{figure}

\textit{Band structure and density of states.}--- Superconducting
cuprates vary from strongly anisotropic to nearly isotropic
materials. To account for this, we consider two limiting cases:
two-dimensional square and 3D cubic symmetries. The obtained
results are quite similar and below we present the band structure
and DOS (Figs.~1 and 2) calculated using Eqs.~\eqref{es} and
\eqref{rhoS0} only for the case of simple cubic lattice. Two very
different quasiparticle spectra are shown in the insets of
Figs.~1a,b. The anticrossing of two bands shown in the inset in
Fig.~1a corresponds to a metallic behavior for any doping. When
doping increases, the chemical potential $\mu$ (shown by the
dotted green line) shifts upward: at low doping we have one
metallic band and one empty band; then two metallic bands; by
further increasing the doping, one metallic and one filled band.
For larger interband hybridization, $t^{pd}>\sqrt{t^{pp}t^{dd}}$,
we obtain a transition to insulator at some doping level. Indeed,
for some doping, $\mu$ is located in the gap between zones. Nearby
the anticrossing point of the two bands, the spectrum
$\varepsilon_j(\zeta(\textbf{k}))$ becomes flatter. This could
indicate the so-called nesting of the Fermi surface. For energies
close to anticrossing points, the DOS exhibit peaks (Fig.~1),
which are large, $\rho_j\propto1/\sqrt{E_0-E}$, when $n$ is close
to 1 in the vicinity of the metal-insulator transition (Fig.~1b).
The optimal doping for superconductivity corresponds to the case
when $\mu$ is close to the energy where the peak of the DOS is
observed.

\textit{Phase separation in cuprates.}--- The energy,
$E_{\text{hom}}(n)=2\sum_{j}\int_{\mu_{\text{min}}}^{\mu}dE\,E\rho_j(E)$,
of the homogeneous state, versus doping $n$, is shown in the inset
of Fig.~2a. The curvature of $E_{\text{hom}}(n)$ is negative
between the two marked points $n_1$ and $n_2$. This indicates the
instability of the homogeneous state with respect to the
separation into two phases with hole densities $n_1$ and $n_2$,
$n_1<n<n_2$ (see, e.g., Ref.~\onlinecite{Our}). The energy
$E_{\textrm{ps}}(n)=cE_{\textrm{hom}}(n_1)+(1-c)E_{\textrm{hom}}(n_2)$
of the PS state with relative phase fractions $c$ and $1-c$, which
are determined by the charge conservation condition
$n=cn_1+(1-c)n_2$, is lower than $E_{\textrm{hom}}(n)$ between
$n_1$ and $n_2$ (see dashed line in the inset of Fig.~2). Thus, in
a wide parameter range, the hole concentration $n_2$ occurs near
the optimum value. A typical dependence of the phase concentration
$c$ versus doping is shown in the inset of Fig.~2b. For low
doping, the system is in a homogeneous state ($c=1$). If $n>n_1$,
the sample divides into droplets with two different hole
concentrations, $n_1$ and $n_2$. Further increasing $n$, the
relative concentration $c$ of the phase with lower hole content
$n_1$ decreases almost linearly, and when $n>n_2$ this phase
disappears and the system becomes homogeneous again. Further
analysis shows that the PS state occurs in the parameter range
where $w_{pd}<w_{pp}$. If $w_{pd}\gg w_{pp}$, the two-band Hubbard
Hamiltonian reduces to an effective single-band model, as is
commonly accepted~\cite{book,ZR} and the discussed cause for PS in
the system disappears.

The PS leads to a redistribution of the charge carriers and charge
neutrality breaking. The structure of the PS state can be either
ordered (e.g., checkerboard structure) or random (i.e., randomly
distributed droplets of one phase within a matrix of the other
phase). For the random PS state, the size ($D$) of the droplet is
determined by the competition between the Coulomb and surface
energies. Following the approach developed in
Ref.~\onlinecite{Our}, we can estimate $D$ as $(3-6)d$ at
$w_{pd}\sim w_{pp} \sim1$~eV, in agreement with the experimental
data in~\cite{droplets}, where spatial variations of the DOS and
superconducting gap were measured using STM.

\textit{Critical temperature versus doping.}--- Hereafter we do
not focus on a precise mechanism of superconductivity in cuprates.
Instead, we intend to analyze the effect of the two-band structure
of the Hubbard Hamiltonian \eqref{H} and PS on the critical
temperature $T_c$ of the superconducting transition. To estimate
$T_c$ we now use the BCS formula~\cite{degen} accounting for the
Coulomb repulsion
\begin{equation}\label{Tc}
T_c=T_D\exp\left[-\;1/\left(\rho_jV_p-\nu_c\right)\right],
\end{equation}
where $T_D$ is the Debye temperature, $V_p$ is the BCS
electron-phonon coupling constant,
$\nu_c=\rho_jV_c/[1+\rho_jV_c\ln(\mu/k_BT_D)]$ is the Coulomb
pseudo-potential, and $V_c\sim U^p\approx 5$~eV are the Coulomb
matrix elements. In our approach, the homogeneous state
superconductivity can appear in different bands depending on
doping, see Fig.~1: in band $j=1$ for low doping and in $j=-1$ for
higher doping. We stress that this is consistent with the
superconducting behavior of YBCO where superconductivity switches
from 60~K to 90~K whith increasing the oxygen content. For
intermediate doping, the superconducting state can simultaneously
coexist in two bands, as in MgB$_2$. Our model could also be used
to interpret experiments~\cite{kunchur} showing the existence of
PS in MgB$_2$. However, a direct application of our studies for
MgB$_2$ should be justified since it is not clear whether the
Hubbard-like description is applicable to this compound.

The dependence of $T_c$ on doping $n$ for the homogeneous state is
shown by the dashed line in Fig.~2b. For the optimum doping level,
corresponding to the anticrossing point (see insets in Fig.~1) of
two hole energies, $T_c$ reaches its maximum because of the peak
in the DOS. Away from optimal doping, $T_c(n)$ decreases fast with
$n$. In the PS state, one of the phases retains the optimum hole
concentration within a wide interval of doping levels. This
results in a slower decrease (see blue solid line in Fig.~2b) of
$T_c$ when $n$ deviates from the optimal value $n_2$. This
provides a possible natural explanation of the observed $T_c$
dependence on hole doping in cuprates.

\textit{Conclusions.}--- We studied the two-band Hubbard model for
superconducting cuprates and other multiband superconductors.
Using the Green's functions technique, we calculated the hole band
structure and the density of states at the Fermi level. The
density of states exhibits a large peak at the anticrossing of
hole bands. The peak is due to the flattening of the Fermi
surface. Near this doping level, a transition from metal to
insulator can occur. Due to the peak in the density of states,
$T_c$ can significantly increase. For a certain range of the model
parameters, a spatial phase separation of the hole carriers
between the two Hubbard bands was found. The discussed mechanism
of PS is an alternative to the usual explanation of the PS, which
is attributed to strong antiferromagnetic correlations~(see,
e.g.,~\cite{Nag,Dag}). Our estimates of the spatial scale of PS is
in a good agreement with important experimental
results~\cite{droplets}. Using the BCS expression for $T_c$, also
accounting for Coulomb repulsion, we show that $T_c(n)$ is near
its maximum within the doping range where PS occurs.

We acknowledge stimulating discussions with A.V.~Rozhkov and
partial support from the NSA, LPS, ARO, NSF (grant No.
EIA-0130383), JSPS-RFBR (grant No. 06-02-91200), CTC program
supported by JSPS, RFBR (grants No. 06-02-16691 and 05-02-17600),
MEXT Grant-in-Aid No. 18740224, the EPSRC {grant No.
EP/D072581/1), and EU project CoMePhs.



\begin{references}
\bibitem{hermit} A non-zero average $\langle a^{\dag}_\alpha a_{\bar{\alpha}}\rangle$
is a well-known internal inconsistency of the Hubbard I
approximation at $t_{ab}\neq 0$~\cite{HubbardIII} since it
produces a non-Hermitian behavior of the Green functions. However,
for the parameter range discussed here, these values are small
compared to the coefficients $g_{\alpha}$ that characterize the
maximum band filling in our model, $|\langle a^{\dag}_\alpha
a_{\bar{\alpha}}\rangle|^2\ll g_{\alpha}g_{\bar{\alpha}}$. It can
be shown that under the latter condition the effect of $\langle
a^{\dag}_\alpha a_{\bar{\alpha}}\rangle$ on the band structure and
band filling is negligible.
\end{references}
\end{document}